\documentclass[%
 reprint,
 superscriptaddress,
 amsmath,amssymb,
 aps,
pre,
]{revtex4-2}

\usepackage{color}
\usepackage{graphicx}
\usepackage{dcolumn}
\usepackage{bm}
\usepackage{ulem}
\usepackage[squaren]{SIunits}

\begin{document}

\preprint{APS/123-QED}

\title{On the generation of attosecond gigawatt soft X-ray pulses through coherent Thomson backscattering}

\author{Qianyi Ma}
 \affiliation{State Key Laboratory of Nuclear Physics and Technology, and Key Laboratory of HEDP of the Ministry of Education, CAPT, School of Physics, Peking University, Beijing 100871, China}
\author{Jiaxin Liu}
 \affiliation{State Key Laboratory of Nuclear Physics and Technology, and Key Laboratory of HEDP of the Ministry of Education, CAPT, School of Physics, Peking University, Beijing 100871, China}
\author{Zhuo Pan}
 \affiliation{State Key Laboratory of Nuclear Physics and Technology, and Key Laboratory of HEDP of the Ministry of Education, CAPT, School of Physics, Peking University, Beijing 100871, China}
\author{Xuezhi Wu}
 \affiliation{State Key Laboratory of Nuclear Physics and Technology, and Key Laboratory of HEDP of the Ministry of Education, CAPT, School of Physics, Peking University, Beijing 100871, China}
\author{Huangang Lu}
 \affiliation{State Key Laboratory of Nuclear Physics and Technology, and Key Laboratory of HEDP of the Ministry of Education, CAPT, School of Physics, Peking University, Beijing 100871, China}
\author{Zhenan Wang}
 \affiliation{State Key Laboratory of Nuclear Physics and Technology, and Key Laboratory of HEDP of the Ministry of Education, CAPT, School of Physics, Peking University, Beijing 100871, China}
\author{Yuhui Xia}
 \affiliation{State Key Laboratory of Nuclear Physics and Technology, and Key Laboratory of HEDP of the Ministry of Education, CAPT, School of Physics, Peking University, Beijing 100871, China}
\author{Yuekai Chen}
 \affiliation{State Key Laboratory of Nuclear Physics and Technology, and Key Laboratory of HEDP of the Ministry of Education, CAPT, School of Physics, Peking University, Beijing 100871, China}
 \author{Kyle Miller}
\affiliation{University of Rochester, Laboratory for Laser Energetics, Rochester, NY 14623, USA}
 \author{Xinlu Xu}
\email[]{xuxinlu@pku.edu.cn}
\affiliation{State Key Laboratory of Nuclear Physics and Technology, and Key Laboratory of HEDP of the Ministry of Education, CAPT, School of Physics, Peking University, Beijing 100871, China}
\affiliation{Beijing Laser Acceleration Innovation Center, Huairou, Beijing, 101400, China}
\author{Xueqing Yan}
 \email[]{x.yan@pku.edu.cn}
 \affiliation{State Key Laboratory of Nuclear Physics and Technology, and Key Laboratory of HEDP of the Ministry of Education, CAPT, School of Physics, Peking University, Beijing 100871, China}
 \affiliation{Beijing Laser Acceleration Innovation Center, Huairou, Beijing, 101400, China}
 \affiliation{CICEO, Shanxi University, Taiyuan, Shanxi 030006, China}
 \affiliation{Institute of Guangdong Laser Plasma Technology, Baiyun, Guangzhou, 510540, China}

\date{\today}

\begin{abstract}
Collision between relativistic electron sheets and counter-propagating laser pulses is recognized as a promising way to produce intense attosecond X-rays through coherent Thomson backscattering (TBS). In a double-layer scheme, the electrons in an ultrathin solid foil are first pushed out by an intense laser driver and then interact with the laser reflected off a second foil to form a high-density relativistic electron sheet with vanishing transverse momentum. However, the repulsion between these concentrated electrons can increase the thickness of the layer, reducing both its density and subsequently the coherent TBS. Here, we present a systematic study on the evolution of the flying electron layer and find that its resulting thickness is determined by the interplay between the intrinsic space-charge expansion and the velocity compression induced by the drive laser. How the laser driver, the target areal density, the reflector and the collision laser intensity affect the properties of the produced X-rays is explored. Multi-dimensional particle-in-cell simulations indicate that employing this scheme in the nonlinear regime has the potential to stably produce soft X-rays with several $\giga\watt$ peak power in hundreds of TW ultrafast laser facilities. The pulse duration can be tuned to tens of attoseconds. This compact and intense attosecond X-ray source may have broad applications in attosecond science. 
\end{abstract}

\maketitle

\section{Introduction}
As one of the most useful tools for exploring and controlling processes with atomic timescale resolution, attosecond radiation sources have broad applications in ultrafast chemistry and physics~\cite{attosecond-physics,attosecond-spectroscopy,attosecond-application,attosecond-imaging}. The extension of attosecond pulses to the soft X-ray regime permits tracking atomic-scale electron motion such as the Auger-Meitner process~\cite{as-xray-coherent-electron-AM-decay} and enables the study of attosecond transient absorption spectroscopy~\cite{as-xray-sbsorption,as-xray-pump-probe}. As the major attosecond sources \cite{attosecond-production}, high harmonic generation (HHG) based on laser-gas interaction \cite{gas-hhg} or laser-plasma interaction \cite{hhg-plasma} can deliver attosecond pulses in the extreme ultraviolet regime with peak power of $\sim 10~\giga\watt$ \cite{20-gw-attosecond1,20-gw-attosecond2}. However, the conversion efficiency of HHG sources is very low at the soft X-ray wavelength ($\sim\nano\meter$), typically less than $10^{-6}$~\cite{water-window-and-solid-state,water-window-as}. Relativistic transition radiation from intense beam-plasma interaction has been proposed to produce terawatt attosecond vacuum ultraviolet radiation \cite{xu2021generation}. Free-electron lasers (FELs) provide an alternative way for generating intense attosecond X-ray pulses~\cite{hard-xfel,soft-xfel,emma2021terawatt,xu2023ultra}. Recently, kilometer-long X-ray FELs have produced soft X-ray attosecond pulses with peak power exceeding $100~\giga\watt$ \cite{100gw-xfel}. However, the number of and access to these large-scale facilities are rather limited.

When relativistic electrons collide with a counter-propagating laser pulse, they oscillate under the laser field and emit radiation with a Doppler upshifted frequency, an effect known as Thomson backscattering (TBS). The relativistic Doppler upshifted factor is $D=\frac{1+\beta_x}{1-\beta_x}\approx4\gamma_x^2$~\cite{einstein}, where $\beta_x=\frac{v_x}{c}$ is the speed of the electrons in the direction of collision ($\hat{x}$) and $\gamma_x=\frac{1}{\sqrt{1-\beta_x^2}}$. For example, collision between 5 MeV electrons and an 800 $\nano\meter$ wavelength laser pulse can produce soft X-rays with wavelength $\lambda_\mathrm{r} = 2.1\nano\meter$. If the length of the electrons is much shorter than the radiation wavelength $\lambda_\mathrm{r}$, the radiation emitted from different electrons superimposes coherently, which leads to a dramatic increase of the intensity and narrowing of the spectrum. Such a thin relativistic electron sheet is also referred to as a relativistic electron mirror (REM) ~\cite{kulagin-as-electronbunch,wu-reflector}. The duration of the pulse reflected off the REM is compressed by a factor of $D$ due to the invariance of the oscillation cycles, and thus can easily access the attosecond regime. Coherent TBS is a promising and compact way to produce intense attosecond X-ray pulses. 

\begin{figure}
\includegraphics[keepaspectratio=true,width=76mm]{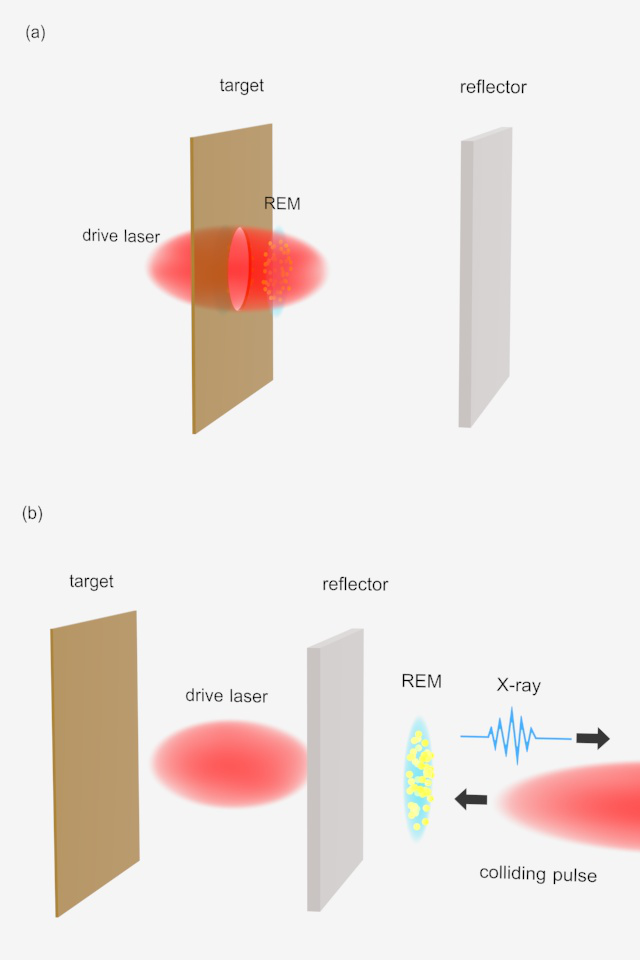}
\caption{\label{fig:1} Schematic of the double-layer scheme for producing a coherent X-ray attosecond pulse. (a) The drive laser irradiates the target and pushes out the electrons to form a relativistic electron mirror (REM). (b) The drive laser is reflected by the reflector, but the REM passes through almost unaffected to reach the colliding pulse and produce an X-ray attosecond pulse. }
\end{figure}

Several schemes have been proposed to generate the needed REMs, such as utilizing the high-density sheath of nonlinear laser wakefield accelerators~\cite{wakefield-sheath-2003,wakefield-sheath-2007,wakefield-sheath-2012} and the electron sheet produced by laser-foil interaction. In the latter, an ultra-short and ultra-intense laser pulse driver is incident normally on a nanofoil. If $a_0 \gg \alpha$, the laser driver pushes out the plasma electrons as a whole \cite{kulagin-as-electronbunch,kulagin-as-electronbunch-feature}, where $a_0 = \frac{eE_0}{m_\mathrm{e} \omega_0 c}$ is the peak normalized vector potential of the laser, $\alpha = \pi \frac{n_0}{n_\mathrm{c}}\frac{d_0}{\lambda_0}$, $e$ and $m_\mathrm{e}$ are the electron charge and mass, $c$ is the speed of light in vacuum, $E_0$ and $\omega_0 = ck_0 = \frac{2\pi c}{\lambda_0}$ are the electric field and frequency of the laser pulse, $n_0$ and $d_0$ are the density and the thickness of the nanofoil, and $n_\mathrm{c}=\frac{m_\mathrm{e}\omega_0^2\varepsilon_0}{e^2}$ is the critical density for the laser pulse. Besides the momentum along the laser propagation direction, however, the electrons also obtain large oscillating transverse momentum, which reduces the longitudinal energy $\gamma_xm_\mathrm{e}c^2$ and brings about a broadband spectrum of the produced radiation~\cite{drawbacks}. An additional reflector has been proposed to be placed behind the nanofoil as shown in Fig. 1 to reflect the drive laser, which then interacts with the electrons to cancel their transverse momentum~\cite{wu-reflector}. In this double-layer scheme, a thin and solid-density REM with approximately uniform $\gamma_x\approx\gamma_0$ can be formed after the reflector, where $\gamma_0$ is the relativistic factor of the electrons. Another counter-propagating laser pulse (the colliding pulse) then collides with the REM to produce an isolated coherent attosecond X-ray pulse as shown in Fig. \ref{fig:1}(b). Some variations of this double-layer scheme have been developed, such as the generation of giant half-cycle attosecond pulses~\cite{wu-half-cycle} and bright subcycle extreme ultraviolet bursts in an oblique double-foil scheme~\cite{ma-subcycle-euv}.

If $d\ll \lambda_\mathrm{r}$, coherent TBS occurs and the intensity of the reflected pulse is proportional to $(nd)^2$, where $d$ and $n$ are the thickness and density of the REM when oscillating in the colliding pulse~\cite{wu-reflectivity}. On the other hand, the strong space-charge force between these highly concentrated electrons leads to a fast expansion of the REM. If $d\gg\lambda_\mathrm{r}$, the TBS process is incoherent and the radiation intensity is proportional to $nd$. Thus, correctly choosing the thickness and density of the nanofoil and understanding its evolution are critical to the generation of an intense attosecond X-ray pulse. Previous works have studied the electron dynamics in the laser field analytically~\cite{wen-mirror-expansion} and found that the REM can be compressed by ultraintense laser drivers with a sharp rising edge \cite{compression-rr}. A circularly polarized or chirped drive laser \cite{cp-mirror,chirp-mirror} and a few-cycle colliding pulse~\cite{thick-mirror-coherent} have been proposed to suppress the expansion of the REM and enhance the coherent TBS efficiency. 

In this paper, we present a systematic study on the dynamic evolution of the REM in a double-foil scheme and find that the competition between velocity compression and space-charge expansion results in a complicated behavior of the REM. The velocity compression results from a negative velocity chirp imparted by the temporally shaped laser driver, which manifests itself differently (1) during ponderomotive acceleration before the reflector and (2) after the transverse momentum cancellation at the reflector. After understanding the roles of the drive laser, the nanofoil and the colliding laser, we propose to accelerate the REM to a higher energy to suppress its expansion, then employ a relativistic colliding laser to produce high-power attosecond soft X-rays through nonlinear coherent TBS. In Sec. \ref{sec2}, the evolution of the REM for different choices of the drive laser and the nanofoil is studied theoretically and with one-dimensional (1D) particle-in-cell (PIC) simulations. In Sec. \ref{sec3}, we study the produced X-ray properties by increasing the REM energy and colliding it with a relativistic colliding laser. The effect of the large energy spread of the REM on the X-rays is presented along with two-dimensional (2D) PIC simulations. A discussion and summary of our findings is given in Sec. \ref{sec5}. 

\section{The evolution of the REM} \label{sec2}
For convenience, we use normalized units, i.e., time and space coordinates are normalized according to $t'=\omega_0 t$ and $x'=\frac{\omega_0}{c} x$, velocity $v'=\frac{v}{c}$, momentum $p'=\frac{p}{m_\mathrm{e}c}$, density $n'=\frac{n}{n_\mathrm{c}}$, and fields $E'=\frac{eE}{m_\mathrm{e}\omega_0 c}$, $B'=\frac{eB}{m_\mathrm{e}\omega_0}$. In the following, these normalized quantities are used without the prime; all physical quantities in this paper are in normalized units unless otherwise specified. 

First, we review the dependence of the reflected pulse intensity on the REM and TBS in 1D geometry. When the REM collides with a counterpropagating laser pulse, part of the laser pulse is reflected with a frequency upshifted factor $D$. The intensity of the reflected pulse is mainly determined by the total electron number $N$ contained in the REM and its bunching factor $b(k_\mathrm{r})$ \cite{RevModPhys.91.035003}, 
\begin{equation}
    I(k_\mathrm{r})\propto \left[ Nb(k_\mathrm{r}) \right]^2 ,
\end{equation}
where $ b(k_\mathrm{r}) = \frac{|\sum_{j=1}^N \mathrm{exp}(ik_\mathrm{r} x_j)|}{N}$, and $k_\mathrm{r}=\frac{2\pi}{\lambda_\mathrm{r}}$ is the wave number of the reflected pulse. If we assume that the REM has a Gaussian density distribution of $n(x)=\frac{N}{\sqrt{2\pi}d}\mathrm{e}^{-\frac{x^2}{2d^2}}$, then $I(k_\mathrm{r})\propto N^2e^{-k_\mathrm{r}^2d^2 } = 2\pi n_\mathrm{peak}^2 d^2 \mathrm{e}^{-k_\mathrm{r}^2d^2 }$, where $n_\mathrm{peak}$ is the peak density of the REM. When the width of the REM increases beyond $d \gtrsim k_\mathrm{r}^{-1}$, the emitted radiation intensity decreases as $d^2\mathrm{e}^{-k_\mathrm{r}^2d^2 }$. When the REM density increases, the radiation intensity increases as $n_\mathrm{peak}^2$. These relations indicate a high-density and thin REM at the collision point is favorable for intense radiation generation. We will present a systematic analysis on the evolution of the REM and clarify how the laser and the nanofoil parameters affect the REM at the collision position. 

In a double-foil scheme, the evolution of the REM can be divided into two stages. In the first stage, all plasma electrons are pushed out to form the REM and accelerated continuously by the ponderomotive force of the drive laser pulse [see Fig.~\ref{fig:1}(a)]. After passing the reflector, the REM drifts in free space with negligible transverse momentum until interacting with the colliding laser pulse, which is referred to as the second stage [see Fig.~\ref{fig:1}(b)]. In the first stage, the single-particle argument based on the canonical momentum conservation gives $\mathbf{p}_\perp(\tau)=\mathbf{a}(\tau)$ and $p_x (\tau)=\gamma (\tau)-1 =\frac{\left| \mathbf{a}(\tau) \right|^2}{2}$, where $\mathbf{p}_\perp$ and $p_x$ are the transverse and longitudinal momentum of the electron, $\mathbf{a}(\tau)\gg 1$ is the normalized vector potential of the drive laser and $\tau=t-x$ is the longitudinal coordinate in a frame which propagates with velocity $c$. If a linearly polarized (LP) drive laser is used, $\gamma$ of the electron oscillates between $1$ and its local maximum in one laser cycle. As a contrast, $\gamma$ follows the laser's envelope in a circularly polarized (CP) laser. Since a large $\gamma$ can suppress the REM expansion through the relativistic effect, a CP laser driver is used in this work \cite{cp-mirror}.

In the first stage (the ponderomotive acceleration stage), the longitudinal velocity of the electrons is given as $v_x (\tau)=\frac{p_x}{\gamma} = \frac{a(\tau)^2}{2+a(\tau)^2}$. Consider two electrons separated by an axial distance of $\Delta x$. Their longitudinal velocity then differs as 
\begin{align}
\Delta v_{x,\mathrm{p}} \approx -\frac{4a}{(2+a^2)^2}\frac{\partial a}{\partial \tau}\Delta x ,
\end{align}
where $\Delta x \ll 1$ is satisfied for parameters of interest. We can see that a given particle is slower than all particles behind it along the rising edge of the drive laser. This negative velocity chirp, i.e., the head of the REM has a smaller forward velocity than the rear, results in a compression of the REM, and a drive laser with a steep rising edge can enhance this velocity compression \cite{compression-rr}. 

However, the space-charge interaction can induce additional velocity difference between particles. After an electron leaves the immobile plasma ions, the 1D sheet model \cite{PhysRev.113.383} gives the longitudinal space-charge field it experiences at $x$ as $E_x (x) = \int_x^{+\infty} \mathrm{d}x'n(x')$. If there is no sheet crossing during the REM evolution, $E_x$ for an electron stays as a constant whose value is only determined by its initial position inside the foil. The difference of $E_x$ between two electrons is $\Delta E_x = n_0 \Delta x_0$, where $\Delta x_0$ represents the initial separation between these two particles, and a nanofoil with uniform density distribution is used. Thus, the accumulated velocity difference in the first stage due to the space-charge interaction is 
\begin{align}
\Delta v_{x,\mathrm{SC}} \approx \frac{8a^2+8}{(a^2+2)^3} n_0\Delta x_0 t .
\end{align}

\begin{figure}[bp]
\includegraphics[keepaspectratio=true,width=80mm]{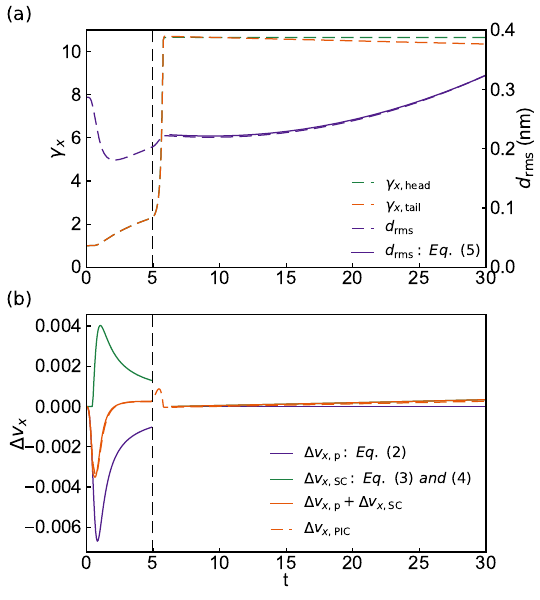}
\caption{\label{fig:2} (a) Evolution of the rms width $d_\mathrm{rms}$ and the forward relativistic factor $\gamma_x$ of the REM head and tail. (b) The velocity difference between the REM head and tail. Solid lines are theoretical results and the dashed lines are from PIC simulation. A low-density and thin foil with $n_0=2,~d_0=0.008~(1~\nano\meter)$ is used and $a_0=10$. A reflector with $n_{0,\mathrm{r}}=1000, ~d_{0,\mathrm{r}}=0.39~(50~\nano\meter)$ is placed at $L=4.0$ away from the nanofoil. The black dashed lines indicate the moment when the REM starts interacting with the reflected drive laser.}
\end{figure}

In the second stage, the transverse momentum of the electrons is canceled by the reflected drive laser and $\gamma_x$ jumps from $\frac{2+a_{0,\mathrm{r}}^2}{2\sqrt{1+a_{0,\mathrm{r}}^2}}$ to $1+\frac{a_{0,\mathrm{r}}^2}{2}$ \cite{wu-reflector}, where $a_{0,\mathrm{r}}$ is the laser's vector potential experienced by the REM at the left boundary of the reflector. During the following drift in vacuum, the space-charge interaction builds a velocity difference between two electrons of 
\begin{align}
\Delta v_{x,\mathrm{SC}} \approx \frac{\Delta E_x (t-t_\mathrm{r})}{\gamma_0^3} = \frac{n_0 \Delta x (t-t_\mathrm{r})}{\gamma_0^3} ,
\end{align}
where $\gamma_0$ is the relativistic factor of the REM in the second stage and $t_\mathrm{r}$ is the time when the transverse momentum of the REM is cancelled. The space-charge force introduces a positive velocity chirp along the REM in both stages, which leads to an expansion. Note that here we take the space-charge force as a small perturbation to the electron's motion, which is valid when the total charge of the nanofoil is low. However, for the parameters we are interested in, the space-charge force may lead to a significant change of the particle's motion. Developing a self-consistent analytical theory is not the goal of this work, and we will rely on PIC simulations to understand the dynamics of the REM.  

\begin{figure*}
\includegraphics[width=1.0\textwidth]{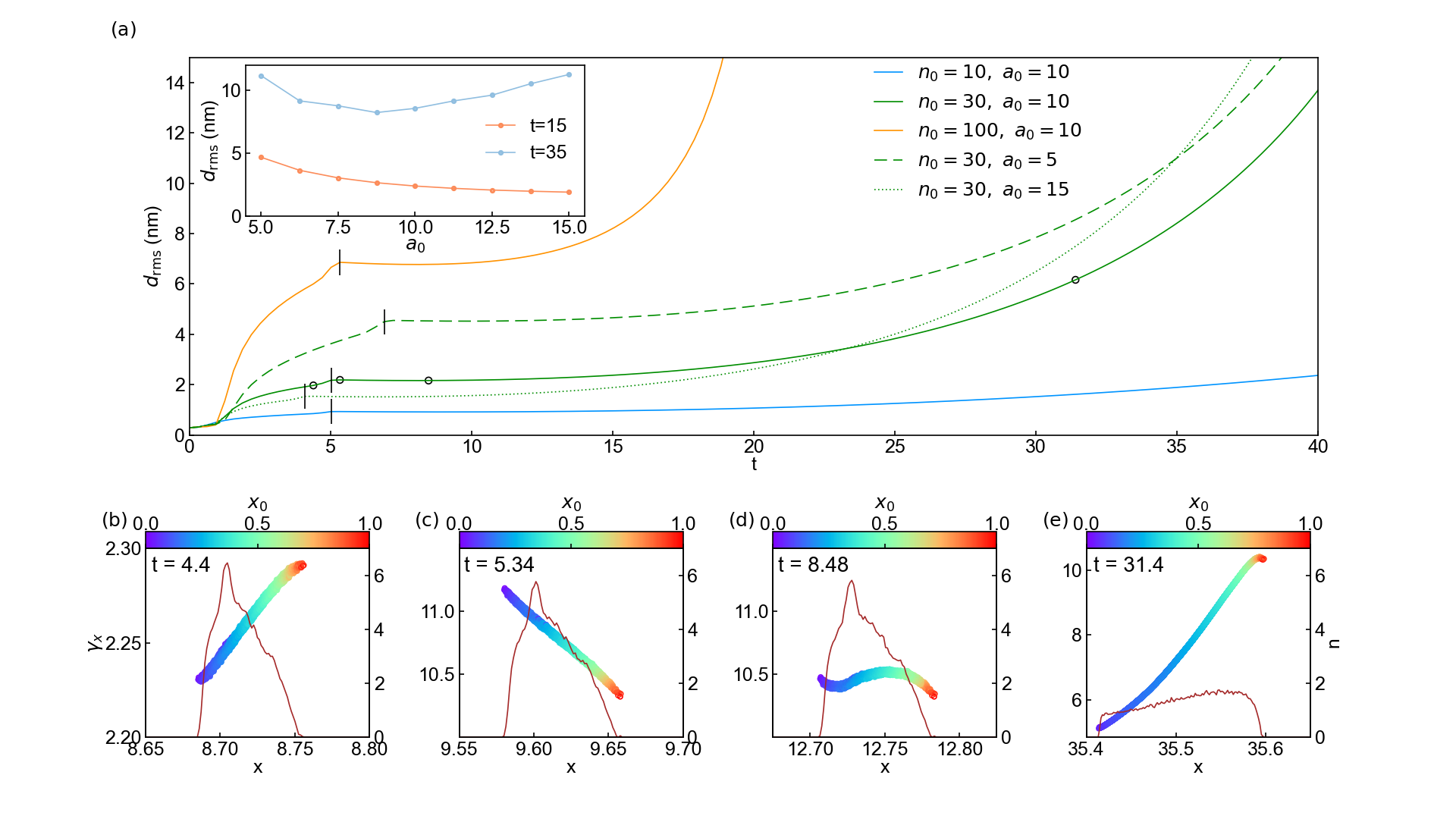}
\caption{\label{fig:3} The evolution of the REM under different $n_0$ and $a_0$. A uniform nanofoil with thickness $d_0=1~\nano\meter$ is used. (a) The change of REM width $d_{\mathrm{rms}}$ for different conditions. The distance between the reflector and the nanofoil is chosen to let the REM drift with $\gamma_{0}\approx 10$, i.e., $L=4.4$ for $a_0=5$, $L=3.1$ for $a_0=10$ and $L=2.4$ for $a_0=15$. The black vertical lines indicate the position of the reflectors. The inset shows the $d_{\mathrm{rms}}$ at $t=15$ and $35$ for different laser intensity when $n_0=30$. (b)-(e) show the longitudinal phase space ($x-\gamma_x$) and the density ($n$) distribution of the REM at different times for $n_0=30$ and $a_0=10$ [indicated by black circles in (a)]. The color represents the position of the particles when they are pushed out of the nanofoil. }
\end{figure*}

To confirm our theoretical analysis and investigate the evolution of the REM, 1D PIC simulations using OSIRIS~\cite{osiris} are performed. A single-cycle CP drive laser pulse with a triangular shape is used for convenient comparison with theory, i.e., $\mathbf{E(\tau)}=\left(0, \frac{a_0}{\pi} \tau \mathrm{cos}\tau, \frac{a_0}{\pi} \tau \mathrm{sin}\tau \right)$ when $0<\tau<\pi$ and $\left(0, \frac{a_0}{\pi} (2\pi-\tau) \mathrm{cos}\tau, \frac{a_0}{\pi} (2\pi - \tau) \mathrm{sin}\tau \right)$ when $\pi \leq \tau<2\pi$. The wavelength of the drive laser is chosen to be $800~\nano\meter$. We use pre-ionized plasmas with a uniform density distribution to represent the nanofoil and the reflector. In a double-layer scheme, the reflector reflects the drive laser to cancel the transverse momentum of the electrons and stop their acceleration. The distance $L$ between the reflector and the nanofoil thus determines the energy of the REM and the corresponding radiation wavelength \cite{tunable-x-ray}. We target 2-$\nano\meter$ radiation and adjust the distance $L$ in the simulations to achieve $\gamma_0 \approx 10$ for the REM. Note that the reflection of the drive laser is not sensitive to the parameters of the reflector if the density and thickness is large enough. A reflectivity of $98.5\%$ is achieved for a drive laser with $a_0=10$ and a reflector with $n_{0,\mathrm{r}}=1000$ and $d_{0,\mathrm{r}}=0.39~ (50~\nano\meter)$ in the simulations. The simulation setup can be found in Appendix A.

We first choose a low-density nanofoil with $n_0=2$ and $d_0=0.008~(1~\nano\meter)$ to examine the above theoretical predictions, taking differences across the entire REM width. A drive laser with $a_0=10$ is incident on the nanofoil at $t=0$ and accelerates the electrons as shown in Fig. \ref{fig:2}(a). For the chosen parameters, the velocity difference $\Delta v_{x,\mathrm{p}}$ due to the ponderomotive force [purple line in Fig. \ref{fig:2}(b)] dominates initially, building a negative chirp and decreasing the root-mean-square (rms) width $d_\mathrm{rms}$ of the REM until $t\approx 2$ [purple line in Fig. \ref{fig:2}(a)]. As $\gamma_x$ increases, the magnitude of both velocity differences decreases due to the relativistic effect. The space-charge term $|\Delta v_{x,\mathrm{SC}}|$ eventually exceeds the ponderomotive acceleration term $|\Delta v_{x,\mathrm{p}}|$ around $t=2.1$, following which the chirp turns from negative to positive and $d_\mathrm{rms}$ starts to increase. At $t=5.0$ (indicated by the black dashed lines), the REM interacts with the reflected drive laser, which cancels its transverse momentum. The $\gamma_x$ jumps quickly from $\sim 2$ to $\sim 10$ and the velocity difference becomes almost zero. During the subsequent drift in vacuum, the space-charge force slowly builds the positive energy/velocity chirp along the REM, and its width increases quadratically:
\begin{align}
d_\mathrm{rms} \approx d_\mathrm{rms,r} + \frac{1}{4\sqrt{3}}\frac{n_0d_0}{\gamma_0^3}(t-t_\mathrm{r})^2 ,
\end{align}
where $d_\mathrm{rms,r}$ is the width when the transverse momentum of the REM is cancelled. A good agreement between this formula (purple solid line) and the PIC result (purple dashed line) is obtained as shown in Fig. \ref{fig:2}(a). Note that we assume the REM has a uniform density distribution in order to calculate its rms width. A good agreement between theory and simulation is obtained in Fig. \ref{fig:2} because a low-density and thin foil is chosen. It is very challenging or impossible to fabricate such low-density and thin foils with current technology. We now consider more realistic parameters.

Figure \ref{fig:3}(a) shows the evolution of the REM width when $n_0=10,~30$ and $100$ and $d_0$ is fixed at 0.008 ($1~\nano\meter$). The drive laser has a more-realistic polynomial envelope with a full width at half maximum (FWHM) duration of $\pi$, and an initial 10-eV temperature is set for the nanofoil electrons (see details in Appendix A). As the nanofoil density becomes large, the velocity difference due to the space-charge force dominates at the beginning and the compression of the REM in the first stage is absent [see Fig. \ref{fig:3}(a)]. We examine the case with $n_0=30$ and $a_0=10$ (green line) and explain its evolution. Before meeting the reflected drive laser ($t=4.4$), a positive energy chirp ($\frac{\mathrm{d} \gamma_x}{\mathrm{d} x} \approx 0.007~\nano\meter^{-1}$) is formed inside the REM by the space-charge interaction as shown in Fig. \ref{fig:3}(b), and the rms width increases from $0.29~\nano\meter$ to $2.2~\nano\meter$. The head and tail of the REM then experience substantially differing values of $a_{0,\mathrm{r}}$ when they reach the reflector, resulting in the negative energy chirp ($\frac{\mathrm{d}\gamma_x}{\mathrm{d} x} \approx 0.086~\nano\meter^{-1}$) shown in Fig. 3(c). This effect is small for the case studied in Fig. \ref{fig:2} since its width at the reflector is 10 times smaller. The space-charge force then gradually turns the negative chirp to positive. At $t=8.5$, the linear energy chirp of the REM is almost removed, while some nonlinear chirps remain as shown in Fig. \ref{fig:3}(d). The chirp then continuously increases, and the growth of the REM width approximately follows a quadratic curve. At $t=31.4$ [Fig. \ref{fig:3}(e)], the REM energy is distributed between $\sim5$ and $\sim10$ with $\frac{\mathrm{d} \gamma_x}{\mathrm{d} x} \approx 0.226~\nano\meter^{-1}$ and $d_\mathrm{rms}\approx 6.16~\nano\meter$. Such a large energy spread affects the radiation properties significantly as discussed in Sec. \ref{sec3}. The color in Fig. \ref{fig:3}(b)-(e) represents the position $x_0$ of the electrons when they are pushed out from the nanofoil. We can see that the assumption of no sheet crossing is approximately valid even with an initial temperature of 10 eV.

\begin{figure}[bp]
\includegraphics[width=0.5\textwidth]{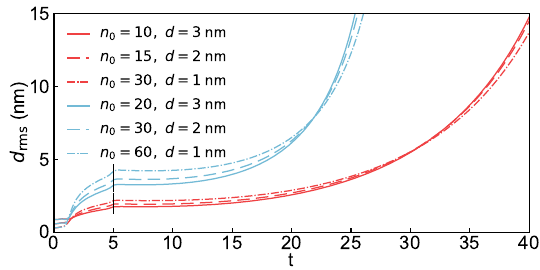}
\caption{\label{fig:n0d0} The evolution of the REM width $d_{rms}$ when varying the nanofoil thickness $d_0$ and density $n_0$. Blue and red lines represent $n_0d_0~[\nano\meter]=30$ and 60, respectively. $a_0=10$ and $L=3.1$ are used to reach $\gamma_0=10$.}
\end{figure}

For the fixed initial width $d_0=1~\nano\meter$, a larger $n_0$ results in a faster growth of $d_\mathrm{rms}$ due to the stronger space-charge interaction [$n_0=100$ indicated by the orange line in Fig. \ref{fig:3}(a)] and vice versa [$n_0=10$ indicated by the blue line in Fig. \ref{fig:3}(a)]. The rms width of the REM for $n_0=100$ increases to $7~\nano\meter$ at the reflector, which is already much longer than the target radiation wavelength of $2~\nano\meter$, after which it continues to grow rapidly. 

The effect of different rising edges of the drive laser is also presented in Fig. \ref{fig:3}(a). A sharp rising edge (equivalent to a large $a_0$ in our simulations) can accelerate the REM quickly and suppress the expansion caused by the space-charge interaction in the first stage. However, the large $d_\mathrm{rms}$ at the reflector for the small $a_0$ case leads to a large negative energy chirp after the reflector, which helps resist broadening from the space-charge interaction and delays the final quadratic growth of $d_\mathrm{rms}$. We show the dependence of $d_\mathrm{rms}$ on $a_0$ at two representative times in the inset of Fig. \ref{fig:3}(a). As $a_0$ increases, $d_\mathrm{rms}$ at $t=15$ decreases because the rapid acceleration suppresses the expansion. As a contrast, there is an optimized $a_0\approx 9$ to achieve a minimum $d_\mathrm{rms}$ at $t=35$, which is due to the competition between the obtained negative chirp after the reflector and the space-charge interaction. 

We can see that the space-charge interaction plays the most critical role in the REM evolution. In 1D geometry, the space-charge interaction is similar for nanofoils with the same areal density $n_0d_0$ as long as $d_0$ varies in a small range. In Fig. \ref{fig:n0d0}, we compare the REM expansion for two groups of parameters: $n_0d_0~[\nano\meter]=30$ and $n_0d_0~[\nano\meter]=60$. As $d_0$ varies from $1~\nano\meter$ to $3~\nano\meter$, the variation of the REM width after the reflector is $<8\%$.

\section{High-power coherent attosecond X-ray pulse generation} \label{sec3}
\begin{figure*}
\includegraphics[keepaspectratio=true,width=160mm]{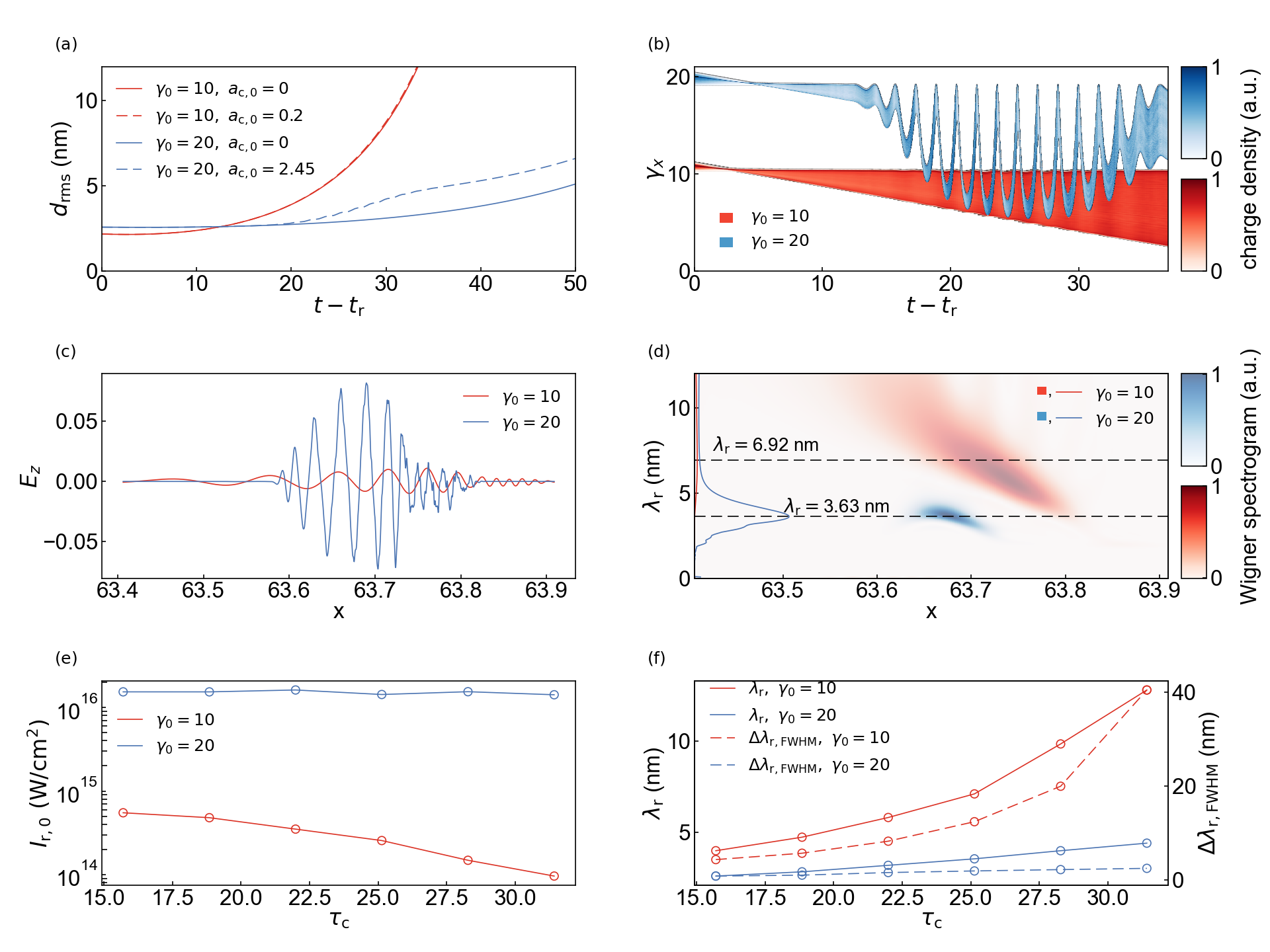}
\caption{\label{fig:5}  Generation of intense attosecond X-ray pulses. (a) The evolution of the REM width $d_\mathrm{rms}$ with and without the colliding pulses. (b) The $\gamma_x$ distribution of the REMs with the colliding pulses. The distribution is normalized at each time for each case. (c) The transverse electric field $E_z$ of the radiation pulses. (d) The normalized Wigner spectrogram of the radiation pulses. The solid lines are the projection of the Wigner spectrogram and the dashed lines indicate their central wavelength. (e) The peak intensity $I_{\mathrm{r},0}$ and (f) the central wavelength $\lambda_\mathrm{r}$ and FWHM bandwidth $\Delta \lambda_{\mathrm{r,FWHM}}$ of the radiation when varying $\tau_\mathrm{c}$.}
\end{figure*}

In Sec. II, we study the evolution of the REM and clarify the role of the ponderomotive acceleration, the reflector and the space-charge interaction. For the parameters we are interested in ($n_0 d_0~[\nano\meter] = 10 \sim 100$), the space-charge expansion dominates the REM evolution and leads to a quadratic growth of $d_\mathrm{rms}$ in the second stage as $\Delta d_\mathrm{rms} \propto \frac{n_0d_0}{\gamma_0^3}$. Here, we propose to accelerate the REM to an energy higher than the required $\gamma_0\approx 10$ to suppress the expansion, then collide it with a relativistic colliding laser pulse to produce the radiation with the target wavelength via nonlinear TBS. The Doppler upshifted factor in the nonlinear regime is $D_\mathrm{NL}=\frac{1}{4\gamma_0^2}\left( 1 + \frac{a_\mathrm{c}^2}{2}\right)$ if an LP colliding laser is used, where $a_c$ is the normalized vector potential of the colliding laser. Nonlinear coherent TBS has been studied by assuming a high-density and ultrathin REM \cite{as-pair} or accelerating the REM to very large energy ($\gamma_0=70$) with a superintense drive laser ($a_0=40$) \cite{wu-nonlinearICS}. The effects of the growth of the energy spread and width on the produced radiation have not been discussed. In this section, we investigate the generation of attosecond X-ray pulses from evolving REMs produced from a double layer scheme in both linear and nonlinear regimes and find that the nonlinear regime can deliver radiation pulses with higher intensity, less wavelength deviation from the designed value and narrower bandwidth as compared with the linear regime. Furthermore, the nonlinear regime can tolerate large timing jitter between the drive laser and the colliding laser pulse.

We move the reflector to $L=6.8$ for the case with $n_0=30,~d_0=1~\nano\meter$ and $a_0=10$ to increase the energy of the REM after the reflector to $\gamma_0\approx 20$. The colliding laser pulse has an 800-$\nano\meter$ wavelength and a 10-cycle polynomial envelope. We choose $a_\mathrm{c,0}=0.2$ for the linear TBS and $a_\mathrm{c,0}=2.45$ for the nonlinear TBS to satisfy $\frac{\lambda_\mathrm{c}}{4\gamma_0^2} \left( 1 + \frac{a_\mathrm{c,0}^2}{2}\right)\approx 2~\nano\meter$, where $a_\mathrm{c,0}$ is the peak normalized vector potential of the colliding laser, and 2 $\nano\meter$ is the target wavelength. The moment when the REM interacts with the peak of the colliding pulse is defined as $t_\mathrm{c}$, and we choose $\tau_\mathrm{c}=t_\mathrm{c}-t_\mathrm{r}=25.1$. Details of the simulations can be found in Appendix A. As shown by the solid lines in Fig. \ref{fig:5}(a), the growth of $d_\mathrm{rms}$ when the REM drifts in the second stage is strongly suppressed by increasing the energy. If a colliding laser pulse is present, the longitudinal velocity of an electron decreases due to the transverse oscillation, i.e., $v_z \approx 1 - \frac{1}{2\gamma_0^2}\left( 1+\frac{a_\mathrm{c}^2}{2}\right)$. This effect enlarges the velocity difference between electrons with different energies when a relativistic colliding laser is used. The dashed lines in Fig. \ref{fig:5}(a) show that the nonlinear TBS enhances the REM expansion (blue) while the linear TBS affects it little (red).

The longitudinal energy $\gamma_x$ distribution of the REM in the second stage is shown in Fig. \ref{fig:5}(b). As described in Sec. II, the energy spread reaches a minimum quickly and then grows continuously. The absolute energy spread for $\gamma_0\approx 10$ and 20 is similar before interacting with the colliding laser for the same $n_0d_0$. The relativistic colliding laser pulse imparts large transverse momentum to the REM electrons and decreases their $\gamma_x$ significantly. 

The temporal distribution of the produced radiation for both cases is presented in Fig. \ref{fig:5}(c). The peak radiation field is $E_z\approx 0.01$ for the linear TBS (red line) and $E_z=0.08$ for the nonlinear TBS (blue line). Besides the suppression of the REM expansion by increasing its energy, the nonlinear TBS itself also contributes to the improvement of the radiation intensity \cite{wu-nonlinearICS}. As a contrast to the sinusoidal profile of the radiation field in the linear TBS, a triangle profile is present in the nonlinear TBS due to the production of harmonics. The pulses in both cases have a similar number of cycles as the colliding laser, i.e., 10 cycles. The FWHM duration in the nonlinear case is $38.4~\atto\second$. The duration can be easily tuned by using colliding laser pulses with different cycles. 

The Wigner spectrogram of the radiation is presented in Fig. \ref{fig:5}(d). In the linear case, the energy chirp of the REM electrons is mapped to the radiation spectrum through the Doppler upshifted factor, i.e., $\frac{\Delta \lambda_\mathrm{r}}{\lambda_\mathrm{r}}\approx 2 \frac{\Delta \gamma}{\gamma_0^3}$. A negative wavelength chirp of $0.1~\nano\meter/\atto\second$ is formed for our parameters. Since the large $a_\mathrm{c}$ in the nonlinear case decreases the relative spread of $\gamma_x$ as shown in Fig. \ref{fig:5}(b), the resulting radiation has a narrow bandwidth. The polynomial envelope of the colliding laser leads to a curved temporal-spectral distribution as shown in Fig. \ref{fig:5}(d) \cite{wu-nonlinearICS}. Due to the continuous decrease of average energy of the REM, the central wavelength of the radiation is always longer than the designed value, which is calculated using $\gamma_0=10$ ($\gamma_0=20$) for the linear (nonlinear) case. Since the wavelength deviation $\Delta\lambda_\mathrm{r}$ is proportional to $\gamma_0^{-3}$, the case with $\gamma_0\approx 20$ has a much smaller deviation than the $\gamma=10$ case.  As shown in Fig. \ref{fig:5}(d), the central wavelength is $\lambda_\mathrm{r}\approx 6.92~\nano\meter$ ($\lambda_\mathrm{r}\approx 3.63~\nano\meter$) for the linear (nonlinear) case.

Since the REM expansion and $\gamma_x$ spread are suppressed for the $\gamma=20$ case, a large change of $\tau_\mathrm{c}$ can be tolerated. In Figs. \ref{fig:5}(e) and (f), the dependence of the intensity, central wavelength and the bandwidth of the produced radiation on $\tau_\mathrm{c}$ is compared for the two cases. We can see that as $\tau_\mathrm{c}$ changes from 15 to 32, the radiation intensity for the linear case decreases by one order of magnitude but remains relatively constant for the nonlinear case. The wavelength and the bandwidth increase in both cases with increasing $\tau_\mathrm{c}$, but are much less sensitive to $\tau_\mathrm{c}$ in the nonlinear case than in the linear case. Thus, the nonlinear regime looses the requirement of the synchronization between the drive laser and the colliding laser pulses.

\begin{figure*}
\includegraphics[keepaspectratio=true,width=160mm]{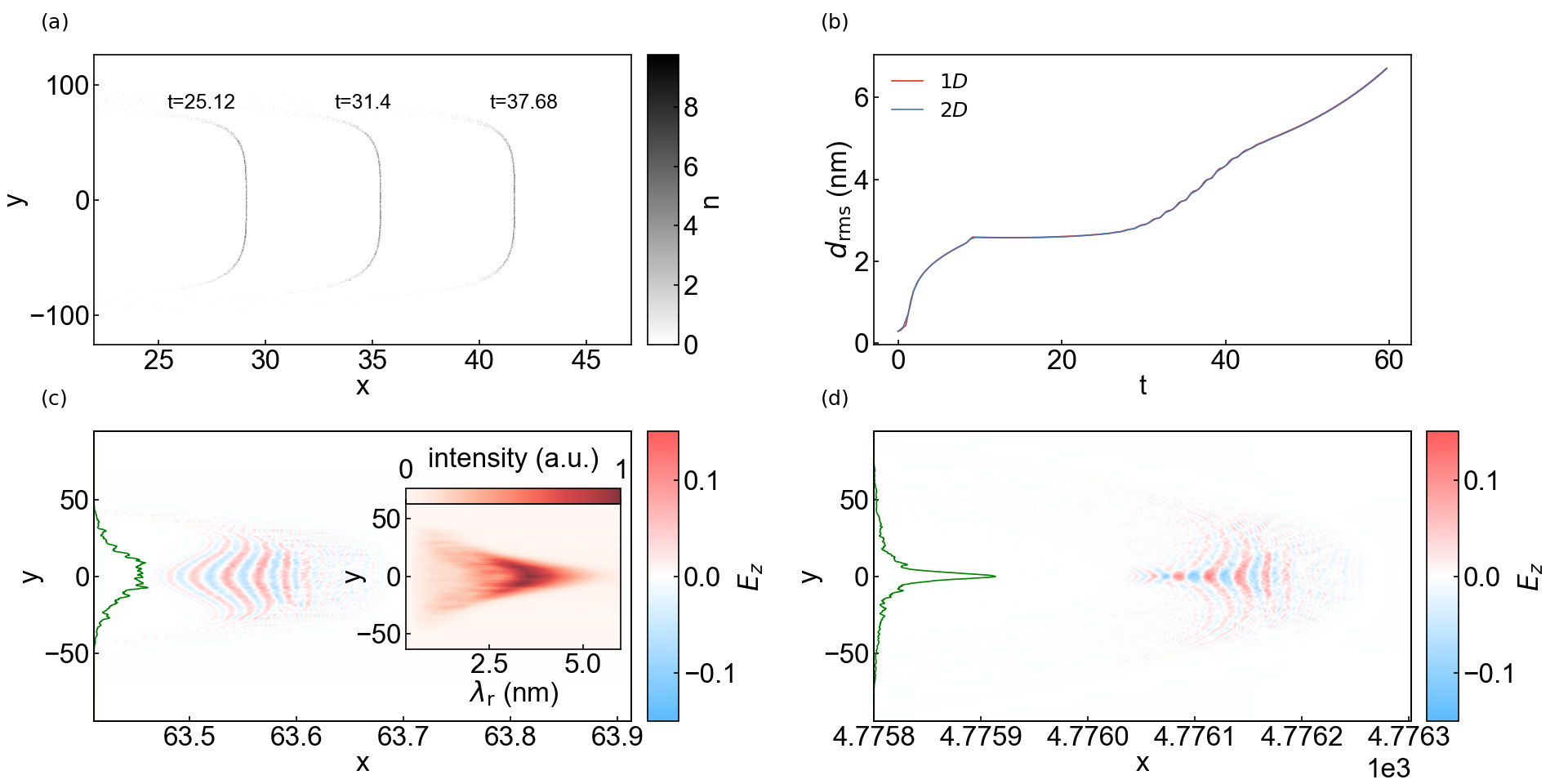}
\caption{\label{fig:6} 2D simulations of intense attosecond X-ray pulse generation. (a) The charge-density distribution at $t=25.12, 31.4$ and $37.68$ without the colliding laser pulse. (b) The comparison of $d_\mathrm{rms}$ between 1D and 2D simulations. Note $d_\mathrm{rms}$ in 2D represents the on-axis thickness. (c) The distribution of the radiation field $E_z$. The inset shows the spectrum along the $\hat{y}$-direction. (d) The distribution of the radiation field after $0.6~\milli\meter$ of propagation. The green lines in (c) and (d) are the longitudinally integrated lineouts of the intensity.}
\end{figure*}

2D PIC simulations have been carried out to study possible multi-dimensional effects. The transverse profile of the drive laser is chosen as a super-Gaussian distribution of $\mathrm{exp}(-\frac{y^4}{R^4})$, where $R=10\lambda_0$. A colliding laser with Gaussian transverse profile is used and the spot size is chosen as $w_0=5\lambda_0$ to ensure the majority is reflected by the flat region of the REM. Simulation setup can be found in Appendix A. The charge density distribution of the REM at several times ($t=25.12,~31.4,~37.68$) is shown in Fig. \ref{fig:6}(a). The REM maintains a quasi-1D density distribution during its drift. The evolution of the on-axis REM width $d_\mathrm{rms}$ is very similar to the 1D result, as shown in Fig.~\ref{fig:6}(b). Fig.~\ref{fig:6}(c) shows the electric field of the produced attosecond X-ray pulse at $t=60$. The Gaussian transverse distribution of the colliding laser leads to two consequences: one is a wavelength change along the $\hat{y}$-direction as shown in the inset of (c) and the other is a focusing wavefront as shown in (c) due to the large $\gamma_x$ of the outermost electrons inside the colliding laser. After $0.6~\milli\meter$ of propagation, the FWHM spot size of the pulse is reduced by a factor of 5, from $3.63~\mu\meter$ to $0.77~\mu\meter$, and its distribution is shown in Fig. \ref{fig:6}(d). 

By assuming the laser pulses and the radiation are axisymmetric, a 270 TW drive laser pulse and a 2 TW colliding laser pulse are used in the above simulation to produce a soft X-ray pulse with 48 as duration and 3 GW peak power. The spot size of the attosecond pulse can be adjusted by employing drive lasers and colliding lasers with different spot sizes, and its pulse duration can be adjusted by changing the duration of the colliding pulse.

\section{Discussion and conclusion} \label{sec5}
One of the biggest challenges involved in this work is the nanofoil fabrication. Promising candidates are nanomeshed graphene~\cite{nm-graphene} and diamond-like carbon foil~\cite{nm-dlc}. The scattering of the REM electrons in the reflector is not included in the PIC simulations, but this effect is unimportant for parameters of interest. According to the classical multiple-scattering formula~\cite{scattering-angle}, assuming we use the Au reflector, the REM will own a scattering angle of a few milliradians after passing through the reflector in our simulations, which has little effect on the subsequent TBS process.

In conclusion, we investigate the generation of intense attosecond X-ray pulses through coherent TBS in a double-foil scheme. The competition between the velocity compression before and after the reflector and the space-charge interaction determines the evolution of the REM. We clarify how the laser pulse driver, the nanofoil and the reflector affect the REM. As a contrast to previous studies of REM, the space-charge interaction results in a large energy chirp on the REM, which leads to a wavelength deviation and a large bandwidth of the radiation. We propose to increase the REM energy to suppress its expansion and produce radiation with a target wavelength of $2~\nano\meter$ through nonlinear TBS. We find that this nonlinear regime can deliver X-rays with high intensity, narrow bandwidth and less wavelength deviation and loose the synchronization requirement between the drive laser and the colliding laser. In multi-dimensional scenario, the attosecond X-ray pulse generated through nonlinear TBS can be focused to smaller spot and higher intensity. Such compact soft X-ray sources with tens of attoseconds duration and several GW peak power may be of general interest in attosecond science.

\begin{acknowledgments}
This work was supported by  the National Natural Science Foundation of China (NSFC) (No. 12375147 and No. 11921006), the National Grand Instrument Project (No. 2019YFF01014400), Guangdong Provincial Science and Technology Plan Project (2021B0909050006), Beijing Outstanding Young Scientist project, and the Fundamental Research Funds for the Central Universities, Peking University. The simulations were supported by the High-performance Computing Platform of Peking University and Tianhe new generation supercomputer at National Supercomputer Center in Tianjin.
\end{acknowledgments}

\appendix
\section{Simulation setup }
For the 1D simulations shown in Sec. \ref{sec2} and Sec. \ref{sec3}, we use a fixed simulation window with a box size of $25\lambda_0$. We choose the grid size $\mathrm{d}x=\frac{\lambda_0}{20000}=0.04~\nano\meter$ to resolve the radiation wavelength ($\sim 2~\nano\meter$) with a time step of $\mathrm{d}t=0.999\frac{\mathrm{d}x}{c}\approx 0.13~\atto\second$ in order to satisfy the Courant-Friedrichs-Lewy (CFL) condition. We use 500 macroparticles per cell to represent the plasma electrons. The ions are taken as immobile. The initial plasma temperature is $T=0$ in Fig. \ref{fig:2}. For other 1D simulations, plasma electrons in the nanofoil have an initial temperature of 10 eV, and the reflector remains cold. The electric field of the polynomial laser pulses in these simulations has a symmetric temporal profile of $10\tau^3-15\tau^4+6\tau^5$, where $\tau=\frac{\sqrt{2}(t-t_0)}{\tau_\mathrm{FWHM}}$.

The 2D simulation in Fig. \ref{fig:6} uses a fixed window with a box size of $17\lambda_0\times 40\lambda_0$ in the $x-y$ plane, and the grid size is $\mathrm{d}x\times \mathrm{d}y=\frac{\lambda_0}{10000}\times\frac{\lambda_0}{40}=0.08~\nano\meter\times 20~\nano\meter$. The time step is $\mathrm{d}t=0.999\frac{1}{\sqrt{(\frac{c}{\mathrm{d}x})^2+(\frac{c}{\mathrm{d}y})^2}}\approx 0.27~\atto\second$. We use 20 macroparticles per cell to represent the plasma electrons. Other settings are the same as the $\gamma_0=20$ case in the 1D simulation in Fig.~\ref{fig:5}. To model the focus of the produced radiation with reasonable computational cost, the electromagnetic fields from Fig. \ref{fig:6}(c) are exported into a small box using a moving window with size of $\frac{2}{25}\lambda_0 \times 20\lambda_0$ in the $x-y$ plane and grid size of $\mathrm{d}x\times \mathrm{d}y=\frac{\lambda_0}{20000}\times\frac{\lambda_0}{40}=0.04~\nano\meter\times 20~\nano\meter$. The time step is $\mathrm{d}t=0.999\frac{1}{\sqrt{(\frac{c}{\mathrm{d}x})^2+(\frac{c}{\mathrm{d}y})^2}}\approx 0.13~\atto\second$.

\bibliography{ref}

\end{document}